\documentclass[12pt,draftclsnofoot, perreview, onecolumn]{IEEEtran}
\usepackage{bbm}
\usepackage{amsfonts}
\usepackage{amsmath}
\usepackage{mathrsfs}
\usepackage{multirow}
\usepackage{booktabs}
\usepackage{setspace}

\usepackage{amssymb,amscd,amsmath}
\usepackage[all]{xy}
\usepackage{fancyhdr}
\usepackage{subfig}
\usepackage{citesort}
\usepackage{graphicx}

\usepackage{graphicx}

\ifCLASSINFOpdf
\else
\fi

\begin{document}
%\begin{spacing}{1.0}
\title{Joint Detection/Decoding Algorithms for Nonbinary LDPC Codes over ISI Channels}
%{New Constructions of Forney's Concatenated Codes with Fountain Applications}
%\author{Dongju Yu,Haiqiang Chen,  Xiao Ma% <-this % stops a space
%\thanks{A, B, C, D, and E are with the Department
%of Electronics and Communication Engineering, Sun Yat-sen
%University, Guangzhou 510275, China  }}
%

%\author{\normalsize{ Shancheng~ Zhao$^1$, Zhifei~Lu$^1$, Xiao~Ma$^1$, Baoming~Bai$^2$ \\% <-this % stops a space
%$^1$Department of Electronics and Communication Engineering\\
% Sun Yat-sen University, Guangzhou, 510275, China E-mail: maxiao@mail.sysu.edu.cn\\
%$^2$The State Key Lab of ISN \\
% Xidian University, Xi¡¯an, 710071, China E-mail:bmbai@mail.xidian.edu.cn}}
\author{ Shancheng~Zhao, Zhifei~Lu, Xiao~Ma, Baoming~Bai
\thanks{Shancheng~Zhao, Zhifei~Lu and Xiao~Ma are with the Department
of Electronics and Communication Engineering, Sun Yat-sen
University, Guangzhou 510275, China. (E-mail: maxiao@mail.sysu.edu.cn, zhaoday@mail2.sysu.edu.cn)}
\thanks{Baoming~Bai is with State Key Lab.~of ISN, Xidian University, Xi'an 710071, China. (E-mail: bmbai@mail.xidian.edu.cn)}
\thanks{This work was supported by 973 Program~(No.2012CB316100) and NSF~(No.61172082 and No. 60972048) of China.}}
\maketitle

\begin{abstract}
%\boldmath

This paper is concerned with the application of nonbinary low-density parity-check (NB-LDPC) codes to binary input inter-symbol interference~(ISI) channels. Two low-complexity joint detection/decoding algorithms are proposed. One is referred to as {\em max-log-MAP/X-EMS algorithm}, which is implemented by exchanging soft messages between the max-log-MAP detector and the extended min-sum~(EMS) decoder. The max-log-MAP/$X$-EMS algorithm is applicable to general NB-LDPC codes. The other one, referred to as {\em Viterbi/GMLGD algorithm}, is designed in particular for majority-logic decodable NB-LDPC codes. The Viterbi/GMLGD algorithm works in an iterative manner by exchanging hard-decisions between the Viterbi detector and the generalized majority-logic decoder~(GMLGD). As a by-product, a variant of the original EMS algorithm is proposed, which is referred to as $\mu$-EMS algorithm. In the $\mu$-EMS algorithm, the messages are truncated according to an adaptive threshold, resulting in a more efficient algorithm. Simulations results show that the max-log-MAP/$X$-EMS algorithm performs as well as the traditional iterative detection/decoding algorithm based on the BCJR algorithm and the QSPA, but with lower complexity. The complexity can be further reduced for majority-logic decodable NB-LDPC codes by executing the Viterbi/GMLGD algorithm with a performance degradation within one dB. Simulation results also confirm that the $\mu$-EMS algorithm requires lower computational loads than the EMS algorithm with a fixed threshold.
These algorithms provide good candidates for trade-offs between performance and complexity.
\end{abstract}

% \fi
\IEEEpeerreviewmaketitle

\section{Introduction}\label{introduction}
Nonbinary low-density parity-check~(NB-LDPC) codes were first introduced by Gallager based on modulo arithmetic~\cite{Gallager63}. In~\cite{Davey98}, Davey and Mackay presented a class of NB-LDPC codes defined over the finite field $\mathbb{F}_q$. They also introduced a Q-ary sum-product algorithm~(QSPA) for decoding NB-LDPC codes. NB-LDPC codes outperform their binary counterparts when used in the channels with burst errors or combined with higher-order modulations. However, the applications of NB-LDPC codes are limited due to their high decoding complexity. To reduce the decoding complexity, a more efficient QSPA based on fast Fourier transform~(FFT-QSPA) was proposed in~\cite{MacKay99}\cite{Barnault03}. To further reduce the decoding complexity, extended min-sum~(EMS) algorithms  were proposed in~\cite{Declercq07}\cite{Voicila10}. The EMS algorithm in~\cite{Voicila10} was re-described in~\cite{Ma2012} as a reduced-search trellis algorithm, called $M$-EMS algorithm. Also presented in~\cite{Ma2012} are two variants of the $M$-EMS algorithm, called $T$-EMS algorithm and $D$-EMS algorithm, respectively\footnote{All these variants are referred to as $X$-EMS algorithms in~\cite{Ma2012}.}. For majority-logic decodable NB-LDPC codes, different low-complexity decoding algorithms have been proposed~\cite{Wangxuepeng10}\cite{Dayuan10}. Different construction methods of NB-LDPC codes have been proposed in the literature, see~\cite{Hu05,Zeng08,Jiang09} and the references therein.

The inter-symbol interference~(ISI) is a common phenomenon in high-density digital recording systems and wireless communication systems~\cite{Proakis}. Different equalizers have been proposed in the literature~\cite{Forney72,Forney73,Bahl74,Kohno86,Prasad88,Douillard95,Sebald02,Cheng07,Peng10,Salamanca12}.
%In~\cite{Peng10}, a low-complexity equalizer with near optimal performance was proposed based on Markov Chain Monte Carlo.
%A joint equalization algorithm based on Bayesian estimation was proposed in~\cite{Salamanca12}.
Since the invention of the turbo codes~\cite{Berrou93}, the rediscovery of the LDPC codes~\cite{Gallager63}, and most importantly, the success of the applications of turbo principle to equalizations~\cite{Douillard95}\cite{Tuchler02}\cite{Gunther05}, many works have been done to apply turbo principles to coded ISI channels~\cite{Kurkoski02,Yang2010,Ryan98,Souviginer00,Song04,Kavcic03,Colavolpe05},
where binary convolutional codes, turbo codes or LDPC codes are usually considered as the ``outer codes" of the serial concatenated system. However, few works are available for the NB-LDPC coded ISI channels. An example is given in Appendix showing that nonbinary may be more suitable to combat inter-symbol interference.

In this paper, we investigate reduced complexity detection/decoding algorithms for NB-LDPC coded ISI channels. Two low-complexity joint detection/decoding algorithms are proposed. For general NB-LDPC coded ISI channels, we propose the max-log-MAP/$X$-EMS algorithm, in which the detector and the decoder are implemented with the max-log-MAP algorithm and the $X$-EMS algorithm, respectively. In this algorithm, the detector takes as input the {\em soft extrinsic messages} from the decoder and delivers as output the soft extrinsic messages of each coded symbol; the decoder takes as input the messages from the detector and feeds back to the detector the soft extrinsic messages of each coded symbol. Simulations results show that the max-log-MAP/$X$-EMS algorithm performs as well as the traditional iterative detection/decoding algorithm based on the BCJR algorithm and the QSPA, but with reduced complexity. Meanwhile, a variant of the original $T$-EMS algorithm is proposed, which is referred to as $\mu$-EMS. The threshold of the $\mu$-EMS algorithm is adaptive and hence can be matched to channel observation. Simulation results show that the proposed $\mu$-EMS algorithm is more effective than the original $T$-EMS algorithm when applied to coded ISI channels. For majority-logic decodable NB-LDPC coded ISI channels, a further complexity-reduced joint detection/decoding algorithm is proposed, referred to as Viterbi/GMLGD algorithm, which is based on the Viterbi algorithm and the generalized majority-logic decoding~(GMLGD) algorithm~\cite{Dayuan10}. In the Viterbi/GMLGD algorithm, the Viterbi detector takes as input the messages from the decoder and delivers as output the {\em hard-decision sequence}; the decoder takes as input the hard-decision sequence from the detector and feeds back to the detector the estimated messages of each coded symbol. Simulations results show that the Viterbi/GMLGD algorithm suffers from a performance degradation within one dB compared with BCJR/QSPA. These algorithms provide good candidates for trade-offs between performance and complexity.

The organization of this paper is as follows. Section II introduces the considered system model. Also given in Section II is
the quantization algorithm to initialize the detector. The max-log-MAP/$X$-EMS algorithms and the Viterbi/GMLGD algorithm are
described in Section III and Section IV, respectively. Complexity comparisons and simulation results are given in Section V. Section VI concludes this paper.

\section{NB-LDPC Coded ISI Channel}\label{a}
\subsection{NB-LDPC Codes}
 Let $\mathbb{F}_q$ be the finite filed with $q = 2^{m}$ elements. A NB-LDPC code $\mathcal{C}_q[N,K]$ is defined as the null
  space of a sparse nonbinary parity-check matrix $\mathbf{H}= [h_{i,j}]_{M \times N}$, where $h_{i,j} \in \mathbb{F}_{q} $. A vector $\underline{v} = (v_0,v_1,v_2,\cdots v_{q-1})\in \mathbb{F}_q^N $ is a codeword if and only if $\mathbf{H} \underline{v}^T = 0$.
   % The parity-check matrix of a majority-logic decodable nonbinary LDPC code has the property that no two rows (or two columns) have more than one position where they both have nonzero-components. This guarantees that
% the Tanner graph of the code is free of cycle of length 4 and hence has girth of at least 6. Let $\lambda_{min}$ denotes the minimum column weight of $\mathbf{H}$. It can be easily shown that the one step majority-logic-decoding~(OSMLGD) algorithm~\cite{Lin04} guarantees to correct all error patterns with weights no more than$ \lfloor \lambda_{min}/2 \rfloor$ and hence that the minimum Hamming weight of the code is lower bounded by $\lambda_{min}+ 1$. In practice, majority-logic decodable nonbinary LDPC codes with relatively large minimum column weights are preferred.
 For convenience, we define the two index sets as follows:
\begin{center}
  $ \mathcal{N}_i = \{j : 0 \leq j \leq N-1,h_{i,j} \neq 0\}   $ \,   for each row $i$ of $\mathbf{H}$
  \end{center}
and
\begin{center}
  $ \mathcal{M}_j = \{i : 0 \leq i \leq M-1,h_{i,j} \neq 0\}   $ \,     for each column $j$ of $\mathbf{H}$.
\end{center}
\subsection{ISI Channel Model }
The ISI channel of order $L$ is characterized by a polynomial
\begin{equation}\label{isi_equation}
f(D) = f_0 + f_1D + f_2D^2 + \cdots + f_LD^L,
\end{equation}
where the coefficients $f_l \in \mathbb{R}$.
%We assume that
%$$\sum_{l = 0}^{L}|f_l|^2 = 1.$$
Let $ x_t \in \mathcal{X} = \{-1,+1\}$ be the channel input at time $t$. The output signal $y_t$ at time $t$ is statistically determined by
\begin{equation}\label{isi_output_equation}
y_t = \sum_{l = 0}^{L}f_{l}x_{t-l} + w_t,
\end{equation}
where $w_t$ is a sample from a white Gaussian noise with two-sided power spectral density $ \sigma^2 = N_0/2$.
%So the signal-to-noise-ratio(SNR) in decibels is given by $ SNR = 10log_{10}\frac{\sum_{l = 0}^{L}f_l^2}{N_0/2} $.

\subsection{The System Model}
\begin{figure*}
\centering
\includegraphics[scale=0.8]{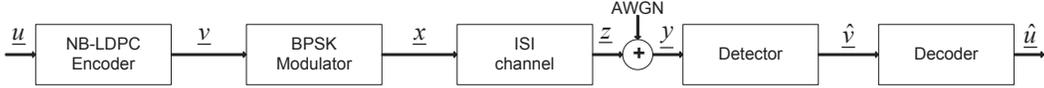}
\caption{System model of a NB-LDPC coded ISI channel.}\label{ISImodel}
\end{figure*}

The system model of a NB-LPDC coded ISI channel is shown in Fig.~\ref{ISImodel}.

{\em Encoding}:
The sequence $\underline{u} = (u_0,u_1,u_2, \cdots ,u_{K-1}) \in \mathbb{F}_q^K$ to be transmitted is encoded by the NB-LDPC encoder into a codeword $\underline{v} = (v_0,v_1,v_2,\cdots,v_{N-1})\in \mathbb{F}_q^N$.

{\em Modulation}:
The codeword $\underline{v}$ is interpreted as a binary sequence $ \underline{c} = (c_0, c_1, \cdots c_n)$ with
$n = mN$ by replacing each component $v_j$ with its binary representation in $\mathbb{F}_q$. The binary sequence $c_j$ is then mapped into a bipolar sequence $\underline{x} = (x_0,x_1,\cdots,x_{n-1})$ with $x_t = 2v_t- 1 $ and transmitted over the ISI channel.

\begin{figure}[!t]
\centering
\includegraphics[scale=0.8]{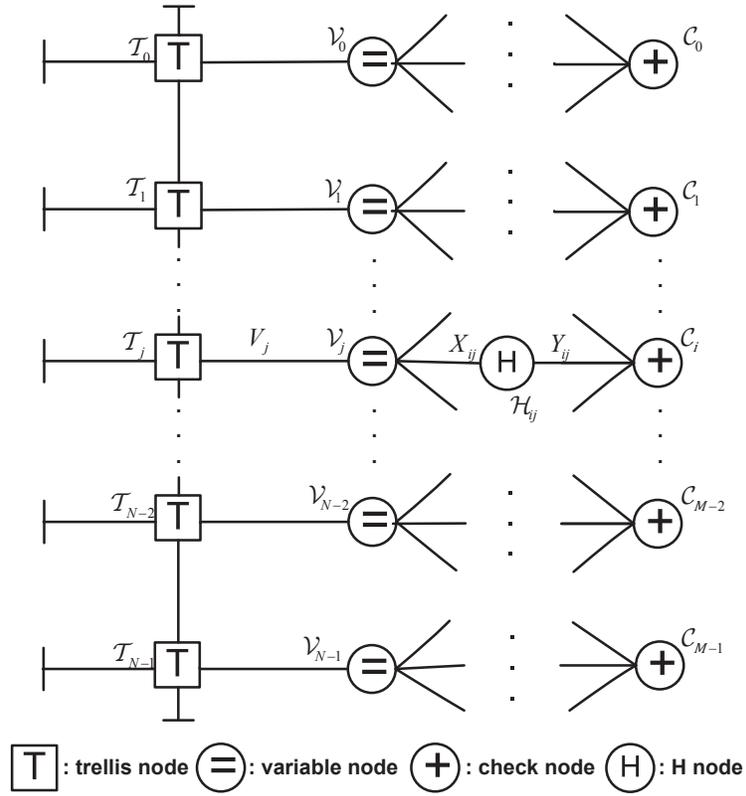}
\caption{Normal graph of a NB-LDPC coded ISI channel. }\label{Normalgraph}
\end{figure}

{\em Detection/Decoding}:
Upon receiving $\underline{y}$, the receiver attempts to recover the transmitted data $\underline{u}$. This can be done by following the well-known turbo principle~\cite{Berrou93} and executing an iterative message processing/passing algorithm~\cite{Kschischang01} over the normal graph~\cite{Forney01} shown in Fig.~\ref{Normalgraph}.
The normal graph has four types of nodes~(constraints): $M$ check nodes~(C-node), $N$ variable nodes~(V-node), $N$ trellis nodes~(T-node) and $\delta$ H-node, where $\delta$ denotes the number of nonzero elements in the parity-check matrix $\mathbf{H}$. The main ingredients of the message processing/passing algorithm include
\begin{itemize}
  \item {\em Detector}: The commonly used detection algorithms are the Viterbi algorithm~\cite{Forney73}, the BCJR algorithm~\cite{Bahl74}\cite{Ma03} and the max-log-MAP algorithm~\cite{Robertson95}\cite{Fossorier98}.
  \item {\em Decoder}: The commonly used decoding algorithm are the QSPA~(or FFT-QSPA)~\cite{Davey98,MacKay99,Barnault03}
  and the $X$-EMS algorithms~\cite{Declercq07}\cite{Ma2012}.
  For majority-logic decodable NB-LDPC codes~\cite{Lin04}, the decoder can also be implemented with the GMLGD algorithm~\cite{Dayuan10}.
\end{itemize}
%\begin{figure}[!t]
%% Requires \usepackage{graphicx}
%\centering
%\includegraphics[scale=1.0]{scheme}
%\caption{Two different schedules for the iterative message passing/processing dection/decoding algorithm.(a)BCJR-once.(b)BCJR-QSPA}
%\end{figure}

Assume that the detector and the decoder are implemented with algorithms {\textbf{A}} and {\textbf{B}}, respectively. We define the following two different schedules
\begin{itemize}
 \item {$\mathbf{A\rightarrow B}$}:~The detector executes the algorithm $\mathbf{A}$ only once, then the decoder performs the decoding algorithm $\mathbf{B}$.
 \item {$\mathbf{A\leftrightarrow B}$}: The detector and the decoder work in an iterative manner by exchanging either soft messages or hard messages between {\textbf{A}} and {\textbf{B}}.
\end{itemize}

In this paper, a reduced complexity detection/decoding algorithm based on the max-log-MAP algorithm
and $X$-EMS algorithms is proposed~(max-log-MAP$ \rightarrow$$X$-EMS or max-log-MAP$\leftrightarrow$$X$-EMS). For majority-logic
decodable NB-LDPC codes, we propose a further reduced complexity detection/decoding algorithm based on the Viterbi
algorithm and the GMLGD algorithm~(Viterbi$ \leftrightarrow $GMLGD). The conventional dectection/decoding algorithms based on the BCJR algorithm and the QSPA, denoted by BCJR$ \rightarrow $QSPA and BCJR$ \leftrightarrow $QSPA, will be taken as
benchmarks for comparison.
\begin{figure}[!t]
%% Requires \usepackage{graphicx}
\centering
\includegraphics[scale=0.65]{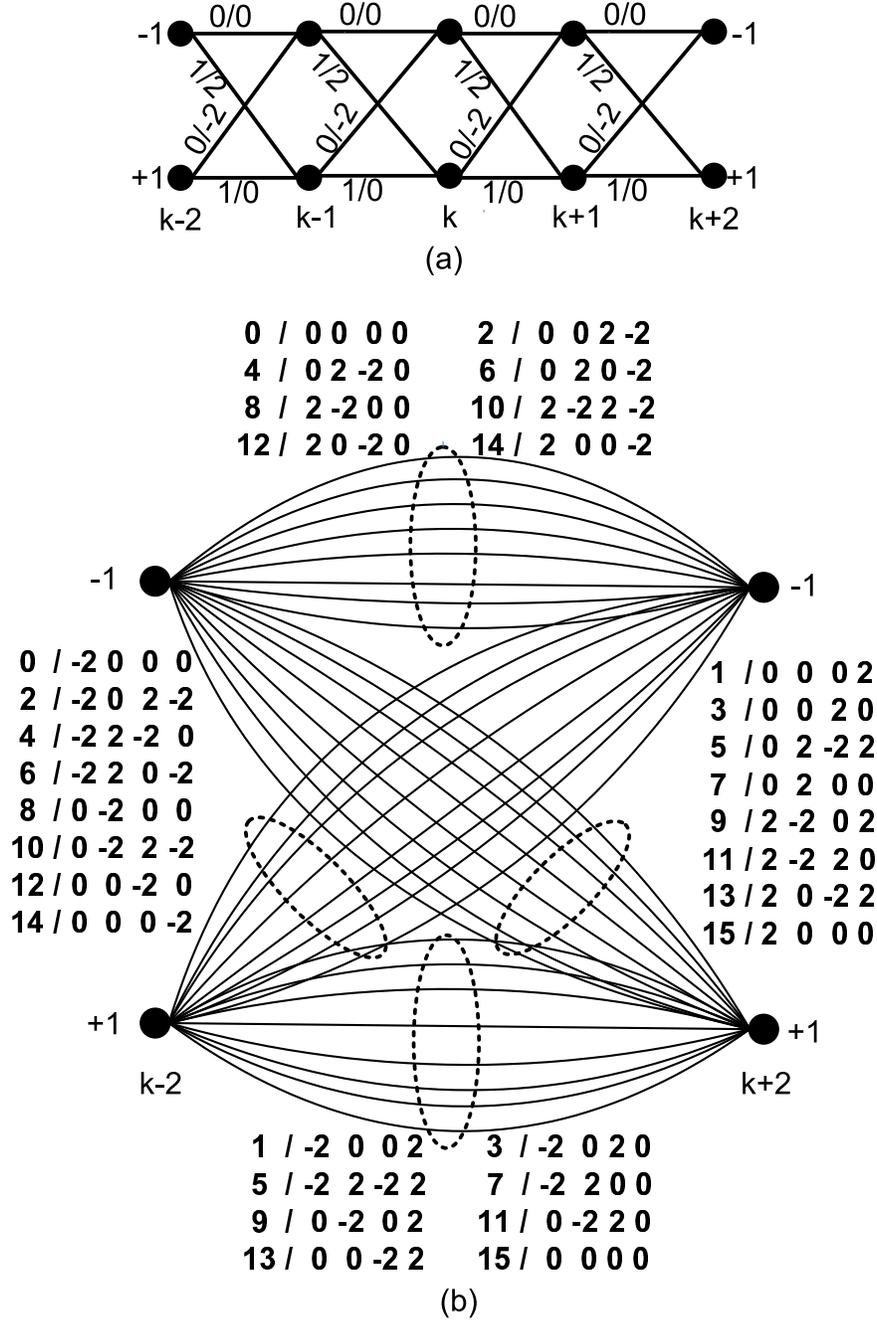}
\caption{Trellis representations of the dicode channel. (a)~The original trellis. (b)~The sectionalized trellis matched to $\mathbb{F}_{16}$.}\label{Trellis}
\end{figure}
\subsection{Sectionalized Trellis}
It has been shown that the ISI channel can be represented by a time-invariant trellis~\cite{Forney72}. At each stage, the trellis has $2^L$ states. Emitting from each state, there are two branches, corresponding to binary inputs 0 and 1, respectively. For convenience, this trellis is referred to as the {\em original} trellis. When an iterative joint decection/decoding algorithm is adopted, we need to exchange messages between the detector and the decoder. The processing of the decoder is symbol-oriented, while the original trellis is bit-oriented. So it is necessary to transform from symbol-based messages to bit-based messages and vice versa, which requires additional computational loads and may cause performance degradations. A way to avoid such a transformation is to work on a $sectionalized$ trellis~\cite{Lafourcade96} directly, which can be obtained from the original trellis. For example, the original trellis and the sectionalized trellis matched to $\mathbb{F}_{16}$ on the the dicode channel $ f(D) = 1 - D $ are illustrated in the Fig.~\ref{Trellis}.
\begin{itemize}
  \item The sectionalized trellis has $N$ section, which are indexed by $ 0 \leq j \leq N - 1$. The $j$-th section corresponds to the $j$-th coded symbol $v_j$.
  \item At each stage, there are $2^L$ states, which are simply indexed by $ 0 \leq s \leq 2^L - 1$. Each state at the $j$-th stage corresponds to a bipolar sequence of the length $L$, that is $s_j \leftrightarrow (x_{(j - 1)L},\cdots,x_{jL - 1} )$, where $x_t$ is the input to the channel at time $t$, and ${x_t, t < 0}$ is assumed to be known at the receiver. The collection of the states at the $j$-th stage is denoted by $ \mathcal{S}_j$.
   \item Emitting from each state, there are $2^m$ branches. Each branch in the $j$-th section is specified by a 4-tuple $ \underline{b} \triangleq (s_j, v_j, {z}_j, s_{j+1})$, where $v_j \in \mathbb{F}_q$ is the $j$-th possible coded symbol that takes the state from $s_j$ into $s_{j+1}$ and results in the noiseless output vector ${z}_j$ of length $m$. In other words, each branch in the sectionalized trellis corresponds to a path of length $m$ in the original trellis. The collection of branches in the $j$-th section is denoted by $\mathcal{B}_j$, we have $|\mathcal{B}_j | = 2^{L + m}$.
\end{itemize}
%Because of using the nonbinary LDPC codes and the sectionalized trellis, the output of the max-log-MAP detector or the Viterbi detector must be a reliability vector $L_{V_j}(z),z \in \mathbb{F}_q $ of the symbol $V_j \in \mathbb{F}_q$.
%
%For conveniently and clearly describing the algorithms, we introduce a basic rule for message processing at an arbitrary node. Let $A$ be a node connecting to $ B_j$ with random variables $ Z_j $ and $Z_j \in \mathbb{F}_q (0\leq j \leq d-1)$, and assume that all incoming messages are available which are denoted by $L_{Z_j}^{(B_j \rightarrow A)}(z), z \in \mathbb{F}_q$. As a message processor, the node $A$ delivers the outgoing message with respect to any given $Z_j$ by computing the
%log likelihood ratio function
%\begin{equation}
% L_{Z_j}^{(B_j \rightarrow A)}(z) = \ln(Prob\{\textrm{A is satisfied }| Z_j = z\})
%\end{equation}
%where the $ L_{Z_j}^{(A \rightarrow B_j )}(z),z\in \mathbb{F}_q$ denote the extrinsic message.
\subsection{Possibility function Calculation }\label{b}
Like most reduced complexity algorithms, we use log-domain messages in the proposed algorithm. Let $Z$ be a discrete random variable taking on values over $\mathcal{Z}$. We use $P_Z(z), z \in \mathcal{Z}$ to denote its probability mass function (pmf). Its $ possibility$ $ function $ is defined as $L_Z(z) =[ a_0\log P_Z(z) + a_1], z\in \mathcal{Z}$, where $[x]$ represents the integer closest to $x\in\mathbb{R}$ and $a_0 > 0, a_1 \in \mathbb{R}$ are two constants. Obviously, we can confine the range of $L_Z(z)$ to be $[0, 2^p-1]$ by properly choosing parameters $a_0$ and $a_1$. In this case $L_Z(z)$ is also referred to as a $p$-bit possibility function. The possibility function can be considered as an integer measure on the possibility of the occurrence of each value $z \in \mathcal{Z}$. Let $X$ denote the variable on the edge connecting the node $\mathcal{A}_1$ and $\mathcal{A}_2$ in
Fig.~\ref{Normalgraph}. We will use $L_{X}^{(\mathcal{A}_1 \rightarrow \mathcal{A}_2)}$ to denote the message
from $\mathcal{A}_1$ to $\mathcal{A}_2$.

To each branch in the $j$-th section of the sectionalized trellis, we assign an integer $L_{Z_{j}}^{(\mid \rightarrow \mathcal{T}_{j})}(z_j)$, where $z_{j}$ is the associated noiseless output. The possibility function $L_{Z_{j}}^{(\mid \rightarrow \mathcal{T}_{j})}(z_j)$ can be determined using the following algorithm.\\
{\bf Algorithm 1:}
Given the received vector $\underline{y}$, $2^p - 1$ and the maximum allowable squared Euclidean distance $d_{max}$ for quantization. For $j=0,1,\cdots,N-1$,
\begin{description}
 \item[{\em Step 1}]: Calculate $d(z_{t})=||y_{j} - z_{j} ||^{2}$, which is the squared Euclidean distance between $y_{j}$ and $z_{j}$;
 \item[{\em Step 2}]: If $d(z_{t}) > d_{max}$, set $d(z_{t}) = d_{max}$;
 \item[{\em Step 3}]: For each noiseless output $z_{j}$, calculate
 \begin{equation}
  L_{Z_{j}}^{(\mid \rightarrow \mathcal{T}_{j})}(z_j)=\left[\frac{d_{max}-d(z_{j})}{d_{max}} \times(2^{p}-1) \right].
 \end{equation}
\end{description}

It can be easily checked that $L_{Z_{j}}^{(\mid \rightarrow \mathcal{T}_{j})}(z_j)$ is a $p$-bit possibility function. For the least possible element $z_{j}$, we have $L_{Z_{j}}^{(\mid \rightarrow \mathcal{T}_{j})}(z_j) = 0$; while for the most possible element $z_{j}$, we have $L_{Z_{j}}^{(\mid \rightarrow \mathcal{T}_{j})}(z_j)\leq 2^{p}-1$. Notice that the variance of the noise is not required to determine $L_{Z_{j}}^{(\mid \rightarrow \mathcal{T}_{j})}(z_j)$.

{\bf Remarks:} It should be pointed out that the maximum allowable Euclidean distance $d_{max}$ is time-invariant which ensures that
$a_0$ in the possibility function is time independent. In this paper, the max-log-MAP algorithm and the Viterbi algorithm are
implemented over the sectionalized trellis with the $p$-bit possibility function as branch metrics. As a result, the detectors require only integer operations.
\section{The Max-log-MAP$\leftrightarrow$$X$-EMS Algorithm}\label{c}
\subsection{T-node: Max-log-MAP Detection}
To each branch $b_{j}=(s_{j},v_{j},z_{j},s_{j+1})$ and $0\leq j <N$, we assign an integer metric
\begin{equation}\label{trellis_equation}
L_{j}(b_{j})= L_{Z_{j}}^{(\mid \rightarrow \mathcal{T}_{j})}(z_j)+ L_{V_j}^{(\mathcal{V}_j \rightarrow \mathcal{T}_{j})}(v_j),
\end{equation}
where $L_{V_j}^{(\mathcal{V}_j \rightarrow \mathcal{T}_{j})}(v_j)$ are initialized as zeros and $L_{Z_{j}}^{(\mid \rightarrow \mathcal{T}_{j})}(z_j)$
is determined by {\bf Algorithm 1}. Then we can execute the max-log-MAP algorithm to obtain an extrinsic possibility vector
$L_{V_j}^{(\mathcal{T}_j \rightarrow \mathcal{V}_{j})}$, for $0\leq j \leq N-1$.

{\bf Remark:} It should be pointed out that the possibility vector $L_{V_j}^{(\mathcal{T}_j \rightarrow \mathcal{V}_{j})}(v_j)$ is normalized such that the reliability of the least possible element is equal to 0.

\subsection{V-node: Computing the Extrinsic Message to H-node}
Given $X_{ij} = x$ , the event of an V-node $\mathcal{V}_{j}$ being satisfied is equivalent to the event $\{V_j = x\}\bigcap_{k\neq i}\{X_{ki} = x\}$. We have
%For a V-node $V_j$, satisfied given $X_{ij} = x$ is equivalent to the event $\{V_j = x\}\bigcap_{k\neq i}\{X_{ki} = x\}$,
\begin{equation}\label{v2h_equation}
L_{X_{ij}}^{(\mathcal{V}_j \rightarrow \mathcal{H}_{ij})}(x) = L_{V_{j}}^{(\mathcal{T}_j \rightarrow \mathcal{V}_{j})}(x) + \sum_{k \neq i} L_{X_{kj}}^{(\mathcal{H}_{kj} \rightarrow \mathcal{V}_{j})}(x),
\end{equation}
where $ L_{V_{j}}^{(\mathcal{T}_j \rightarrow \mathcal{V}_{j})}(x)$ is the message from the max-log-MAP detector and $ L_{X_{kj}}^{(\mathcal{H}_{kj} \rightarrow \mathcal{V}_{j})}(x)$ are initialized as zeros for $x\in \mathbb{F}_q$.

\subsection{H-node: Message Permutation}
Given $Y_{ij} = y$, the event of an H-node $\mathcal{H}_{ij}$ being satisfied is equivalent to the event $\{X_{ij} = h_{ij}^{-1}y\}$. We have
\begin{equation}\label{h2c_equation}
L_{Y_{ij}}^{(\mathcal{H}_{ij} \rightarrow \mathcal{C}_{i})}(y) = L_{X_{ij}}^{(\mathcal{V}_j \rightarrow \mathcal{H}_{ij})}(h_{ij}^{-1}y), \, y\in \mathbb{F}_q.
\end{equation}

\subsection{C-node: Computing the Extrinsic Message to H-node}
{\em Message-truncation rules:}~Given the message $L_{Y_{ij}}^{(\mathcal{H}_{ij} \rightarrow \mathcal{C}_{i})}$, we can partition the finite field $\mathbb{F}_q$
into $\mathcal{F}$ and $\mathbb{F}_q-\mathcal{F}$. Three different message-truncation rules have been proposed in~\cite{Ma2012}. That are

  \begin{eqnarray*}\label{fm_equation}
  \mathcal{F}_M &=&  \{y\in \mathbb{F}_q|L_{Y_{ij}}^{(\mathcal{H}_{ij} \rightarrow \mathcal{C}_{i})}(y)\text{\rm~is one of the }M \text{~largest components of }L_{Y_{ij}}^{(\mathcal{H}_{ij} \rightarrow \mathcal{C}_{i})}\},
  \end{eqnarray*}

  \begin{equation*}\label{ft_equation}
  \mathcal{F}_T = \{y\in \mathbb{F}_q|L_{Y_{ij}}^{(\mathcal{H}_{ij} \rightarrow \mathcal{C}_{i})}(y) \geq T\},
  \end{equation*}
  and
  \begin{equation*}\label{fd_equation}
  \mathcal{F}_D = \{y\in \mathbb{F}_q|L_{max} - L_{Y_{ij}}^{(\mathcal{H}_{ij} \rightarrow \mathcal{C}_{i})}(y)\leq D\},
\end{equation*}
where $L_{max}$ denotes the largest component of $L_{Y_{ij}}^{(\mathcal{H}_{ij} \rightarrow \mathcal{C}_{i})}$ and $D$ is a designated parameter. In this paper, we give a new truncation rule
 \begin{equation*}\label{fmu_equation}
  \mathcal{F}_{\mu} \stackrel{\Delta}{=} \{y\in \mathbb{F}_q|L_{Y_{ij}}^{(\mathcal{H}_{ij}\rightarrow \mathcal{C}_{i})}(y)\geq \mu\},
\end{equation*}
where $\mu$ is determined by
 \begin{equation}\label{mu_equation}
\mu = \frac{1}{q}\sum_{y\in \mathbb{F}_q} L_{Y_{ij}}^{(\mathcal{H}_{ij}\rightarrow \mathcal{C}_{i})}(y) - c,
\end{equation}
where $c$ is a constant to be designated. That is, $\mu$ is equal to the mean of the possibility vector
$L_{Y_{ij}}^{(\mathcal{H}_{ij}\rightarrow \mathcal{C}_{i})}$ with an offset of $c$.
The resultant EMS algorithm is referred to as $\mu$-EMS here.

Given a truncation rule, the possibility vector $L_{Y_{ij}}^{(\mathcal{C}_i \rightarrow \mathcal{H}_{ij})}$ from the C-node $\mathcal{C}_j$ to the H-node $\mathcal{H}_{ij}$ can be calculated by a reduced
trellis search algorithm. See~\cite{Ma2012} for details.

{\bf Remark:}~The truncation rule $\mathcal{F}_{\mu}$ is simpler than the truncation rule $\mathcal{F}_{M}$, since no ordering is required. The truncation rule $\mathcal{F}_{\mu}$ is similar to $\mathcal{F}_{T}$ except that the
threshold of $\mathcal{F}_{\mu}$ is data-dependent and hence can be matched to data and iterations.

\subsection{H-node: Message Permutation}
Given $X_{ij} = x$, the event of an H-node $\mathcal{H}_{ij}$ being satisfied is equivalent to the event $\{Y_{ij} = h_{ij}x\}$. Then
\begin{equation}\label{h2v_equation}
L_{X_{ij}}^{(\mathcal{H}_{ij} \rightarrow \mathcal{V}_{j})}(x) = L_{Y_{ij}}^{(\mathcal{C}_i \rightarrow \mathcal{H}_{ij})}(h_{ij}x), \, x\in \mathbb{F}_q.
\end{equation}
\subsection{V-node: Making Decisions and Computing the Extrinsic Message to T-node}
For the V-node $\mathcal{V}_j$, $0\leq j\leq N-1$, calculate the message
\begin{equation}\label{vd_equation}
L_{V_{j}}(x) = L_{V_{j}}^{(\mathcal{T}_j \rightarrow \mathcal{V}_{j})}(x) + \sum_{i \in \mathcal{M}_j} L_{X_{ij}}^{(\mathcal{H}_{ij} \rightarrow \mathcal{V}_{j})}(x)
\end{equation}
and make decisions according to
\begin{equation}
\hat{v}_j = \arg\max_{x\in \mathbb{F}_q}L_{V_j}(x).
\end{equation}
If $H\hat{\underline{v}}^{T}=\underline{0}$, output $\hat{\underline{v}}$ as the estimated codeword.
If $H\hat{\underline{v}}^{T}\neq \underline{0}$, calculate the message $L_{V_j}^{(\mathcal{V}_j \rightarrow \mathcal{T}_{j})}$ from V-node $\mathcal{V}_j$ to T-node $\mathcal{T}_j$ as
\begin{equation}\label{v2t_equation}
L_{V_j}^{(\mathcal{V}_j \rightarrow \mathcal{T}_{j})}(x) = L_{V_{j}}(x)- L_{V_{j}}^{(\mathcal{T}_j \rightarrow \mathcal{V}_{j})}(x),
\end{equation}
for $x\in \mathbb{F}_q$.
\subsection{Summary of The Max-log-MAP$\leftrightarrow$$X$-EMS Algorithm}
\begin{itemize}
  \item Initialization: Given $\underline{y}$ and a truncation rule $\mathcal{F}$, set a maximum iteration number $\mathcal{L}$ and an iteration variable $l=0$. For all $\mathcal{V}_j$ and $x \in \mathbb{F}_q$, set $L_{V_j}^{(\mathcal{V}_j \rightarrow \mathcal{T}_{j})}(x) = 0$, $ L_{X_{ij}}^{(\mathcal{H}_{ij} \rightarrow \mathcal{V}_{j})}(x) = 0$.
  \item Iteration: while $l< \mathcal{L}:$
  \begin{enumerate}
    \item Detection at T-node: Executing the max-log-MAP algorithm with the branch metrics as defined in~(\ref{trellis_equation}) to obtain
    the possibility vector $ L_{V_j}^{(\mathcal{T}_j \rightarrow \mathcal{V}_{j})}$.
    \item Messages processing at V-node: for all V-nodes, calculate $ L_{X_{ij}}^{(\mathcal{V}_{j} \rightarrow \mathcal{H}_{ij})}$
    according to~(\ref{v2h_equation}).
    \item Messages permutation at H-node: for all H-nodes, permute the messages $L_{Y_{ij}}^{(\mathcal{H}_{ij} \rightarrow \mathcal{C}_{i})}$ according to~(\ref{h2c_equation}).
    \item Messages processing at C-node: for all C-nodes, calculate the messages $L_{Y_{ij}}^{(\mathcal{C}_i \rightarrow \mathcal{H}_{ij})}$ according to the truncation rule $\mathcal{F}$.
    \item Messages permutation at H-node: for all H-nodes, permute the messages $L_{X_{ij}}^{(\mathcal{H}_{ij}\rightarrow \mathcal{V}_{j})}$ according to~(\ref{h2v_equation}).
    \item Messages processing at V-node: for all V-nodes, calculate the messages $L_{V_{j}}$ and find $ \hat{v}_j$.
        If $H\underline{\hat{v}}^T = 0$, output $\underline{\hat{v}}$ and exit the iteration;
        otherwise, calculate the messages $L_{V_j}^{(\mathcal{V}_j \rightarrow \mathcal{T}_{j})}$.
    \item Increment $l$ by one.
  \end{enumerate}
  \item Failure: If $l=\mathcal{L}$, report a decoding failure.
\end{itemize}
{\bf Remark:}~Note that the proposed algorithm requires only integer operations and finite field operations.
\section{The Viterbi$\leftrightarrow$GMLGD Algorithm}\label{d}
For majority-logic decodable NB-LDPC coded ISI channels, we propose a further complexity-reduced joint detection/decoding algorithm
based on the Viterbi algorithm and the GMLGD algorithm. The parity-check matrix of a majority-logic decodable NB-LDPC code~\cite{Lin04} has the property that no two rows (or two columns) have more than one position where they both have nonzero-components. This guarantees that
 the Tanner graph of the code is free of cycle of length 4 and hence has girth of at least 6. In practice, majority-logic decodable NB-LDPC codes with redundant rows~\cite{Huang12} are preferred.
\subsection{T-node: Viterbi Detection}
To each branch $b_{j}=(s_{j},v_{j},z_{j},s_{j+1})$ and $0\leq j <N$, we assign an integer metric
\begin{equation}\label{LV_b}
L_{j}(b_{j})= L_{Z_{j}}^{(\mid \rightarrow \mathcal{T}_{j})}(z_j)+ L_{V_j}^{(\mathcal{V}_j \rightarrow \mathcal{T}_{j})}(v_j),
\end{equation}
where $L_{V_j}^{(\mathcal{V}_j \rightarrow \mathcal{T}_{j})}(v_j)$ are initialized as zeros and $L_{Z_{j}}^{(\mid \rightarrow \mathcal{T}_{j})}(z_j)$
is determined by {\bf Algorithm 1}. Then we can run the Viterbi algorithm through the sectionalized trellis to find a path $\hat{b}_{0},\hat{b}_{1},\cdots,\hat{b}_{N-1}$ such that the path metric $\sum_{0\leq j\leq N-1}L_{j}(\hat{b}_{j})$ is maximized. The associated input sequence $\hat{v}$ is then passed to the variable nodes as the hard decisions.
\subsection{V-node: Syndrome Computation}
After receiving the hard-decision vector $\hat{v}$ from the T-node, we may calculate the syndrome
\begin{equation}\label{s_v}
\underline{s}=\underline{\hat{v}}H^{T}=(s_{0},s_{1},\cdots,s_{m-1}).
\end{equation}
If $\underline{s}=0$, output $\underline{\hat{v}}$ as the decoding result; otherwise, the variable nodes send the hard decision vector $\underline{\hat{v}}$ together with the syndrome vector $\underline{s}$ to the check nodes.
\subsection{C-node: Extrinsic Estimation}
The $i$-th check node sends back an extrinsic estimate to the $j$-th variable node, which is denoted by $\sigma_{i\rightarrow j}$ and can be determined by
\begin{equation}\label{s_c}
\sigma_{i\rightarrow j}=-h^{-1}_{i,j}(\sum_{j'\in \mathcal{N}_{i}\backslash j}h_{i,j'}\hat{v}_{j'})=-h^{-1}_{i,j}s_{j}-\hat{v}_{j},
\end{equation}
where $i\in \mathcal{M}_{j}$ and all the operations are executed in $\mathbb{F}_{q}$.
\subsection{V-node: Possibility Function Updates}
Intuitively, for each variable node $\mathcal{V}_{j}$, the occurrence of each $\alpha \in \mathbb{F}_{q}$ in the received messages $\{\sigma_{i\rightarrow j},i\in \mathcal{M}_j\}$ from check nodes reflects its possibility. Therefore these votes can be used to update the possibility function by increasing $L_{V_j}^{(\mathcal{V}_j \rightarrow \mathcal{T}_{j})}(v_j)$, $v_j\in \mathbb{F}_q$, according to the following rule:
\begin{equation}\label{s_vd}
L_{V_j}^{(\mathcal{V}_j \rightarrow \mathcal{T}_{j})}(\sigma_{i\rightarrow j})\leftarrow L_{V_j}^{(\mathcal{V}_j \rightarrow \mathcal{T}_{j})}(\sigma_{i\rightarrow j})+1,
\end{equation}
for all $ i\in \mathcal{M}_j$. In words, for a given $\alpha \in \mathbb{F}_{q}$, $L_{V_j}^{(\mathcal{V}_j \rightarrow \mathcal{T}_{j})}(\alpha)$ is a counter that accumulates, up to and inclusive of the current iteration, all the occurrences of $V_{j}=v_j$ in the extrinsic messages sent back from the adjacent check nodes.
\subsection{Summary of The Viterbi$\leftrightarrow$GMLGD Algorithm}
\begin{enumerate}
  \item Initialization: Given $\underline{y}$, calculate the $p$-bit possibility functions according to {\bf Algorithm 1}. Select a maximum iteration number $\mathcal{L}>0$ and set $l=0$. For all $\mathcal{V}_{j}$ and $v_j \in \mathbb{F}_{q}$, set $L_{V_j}^{(\mathcal{V}_j \rightarrow \mathcal{T}_{j})}(v_j)=0$.
  \item Iteration: While $l<\mathcal{L}$:
  \begin{enumerate}
    \item Detection at T-node: determines the hard decision sequence $\underline{\hat{v}}$ by executing the Viterbi algorithm with branch metrics as defined in~(\ref{LV_b}).
    \item Syndrome computation at V-node: compute the syndrome $\underline{s}$ according to~(\ref{s_v}).
         If $\underline{s}=0$, output $\underline{\hat{v}}$ and exit the iteration;
         otherwise, send $\underline{s}$ and $\underline{\hat{v}}$ to the C-nodes.
    \item Extrinsic estimation at C-node: compute $\sigma_{i\rightarrow j}$ according to~(\ref{s_c}) and send them to the V-nodes.
    \item Possibility function update at V-node: update the possibility functions $L_{V_j}^{(\mathcal{V}_j \rightarrow \mathcal{T}_{j})}$ according to~(\ref{s_vd}).
    \item Increment $l$ by one;
  \end{enumerate}
  \item Failure Report: If $l=\mathcal{L}$, report a decoding failure.
\end{enumerate}
\section{Complexity Analysis and The Numerical Results}\label{e}
\subsection{Complexity Analysis}
The computational complexities per iteration of the Viterbi algorithm, the BCJR algorithm, the max-log-MAP algorithm, the GMLGD algorithm and the QSPA algorithm are shown in Table~\ref{complexity_table}, where $\delta$ denotes the number of non-zero elements in $\mathbf{H}$. However, the complexity of the $X$-EMS algorithm varies from iteration to iteration.

Apparently, for each iteration, the max-log-MAP$\leftrightarrow$$X$-EMS algorithm and the Viterbi$\leftrightarrow$GMLGD algorithm require less operations than
BCJR$\leftrightarrow$QSPA. However, they may require more iterations to converge. Therefore, for a fair comparison, we take
\begin{equation}\label{complexity_ratio}
\frac{\textrm{total number of operations of a given algorithm }}{\textrm{total number of operations of the BCJR$\leftrightarrow$QSPA algorithm }}
\end{equation}
as the complexity measurement. Note that the statistical mean~(averaging over frames) of the total number of operations involved
in all iterations for decoding one frame is used in (\ref{complexity_ratio}). Also note that the ratio in (\ref{complexity_ratio}) only give a rough comparison, as different algorithms require different operations.
\setlength{\aboverulesep}{0pt}
\setlength{\belowrulesep}{0pt}
\begin{table*}[!t]
 \caption{Computational Complexities of Different Algorithms Required Per Iteration.}\label{complexity_table}
 \center
 \newcommand{\lw} {\vrule width 2pt}
 \begin{tabular}{l||c|c|c||c|l|}
  \toprule
       \multirow{2}{*}{} &\multicolumn{3}{c||}{Detection algorithm}& \multicolumn{2}{c|}{Decoding algorithm} \\ \cline{2-6}
                    & BCJR &Max-log-MAP & Viterbi & QSPA       & GMLGD  \\
  \midrule
  Integer Addition    &        &$4Nq2^L$&$N2^L$   &            &$\delta + Nq2^L$ \\ \hline
  Integer Comparison  &        &$3Nq2^L$&$Nq2^L$  &            &                 \\ \hline
  Field Operation     &        &        &         &$q\delta$   &$4q\delta$       \\ \hline
  Real Multiplication &$4Nq2^L$&        &         &$2q\delta$  &                 \\ \hline
  Real Addition       &$3Nq2^L$&        &         &$2q^2\delta$ &                 \\ \hline
  Real Division       &        &        &         &$2q\delta$  &                 \\
  \bottomrule
 \end{tabular}
\end{table*}

%The simulation results in the Fig.\ref{Complexity_max}show that the max-log-MAP$\leftrightarrow$$X$-EMS algorithms we proposed have a low compute complexity comparing with the BCJR$\leftrightarrow$QSPA algorithm with performs loss ignored, while the Viterbi$\leftrightarrow$GMLGD algorithm proposed have a much lower compute complexity comparing with the BCJR$\leftrightarrow$QSPA algorithm with a little performs loss.

\subsection{Numerical Results}
Let $X_s$ and $X_b$ denote the parameters in the truncation rule $X$ for state metrics and branch metrics, respectively.

\begin{figure}[!t]
    \centering
    \includegraphics[scale=0.9]{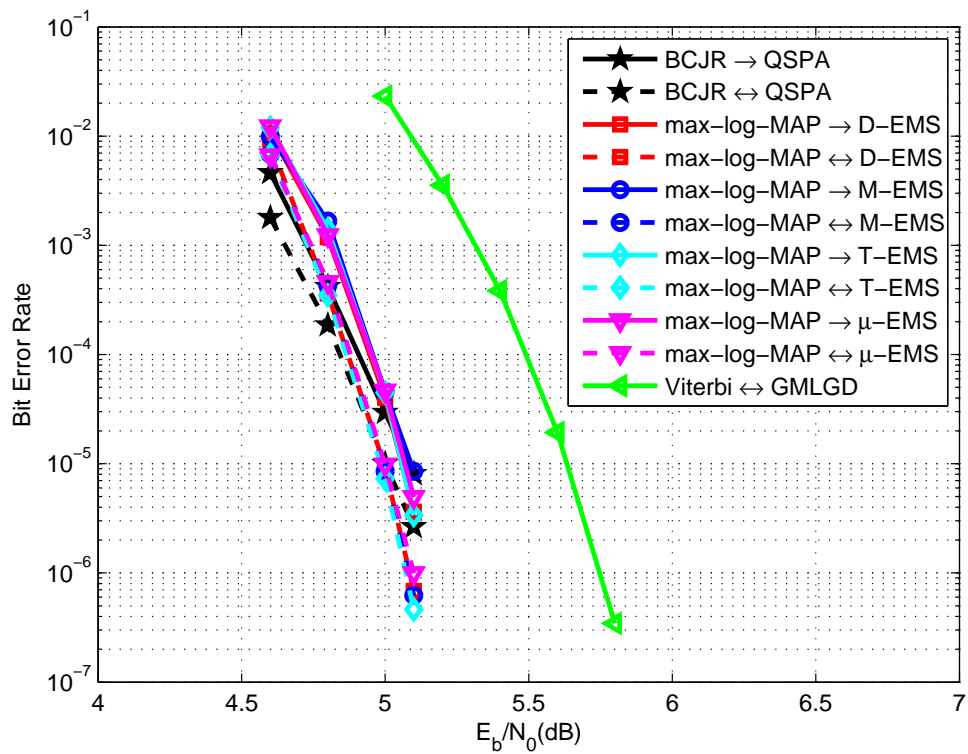}
    \caption{Error performances of different detection/decoding algorithms for decoding the $\mathcal{C}_{32}[961,765]$ coded dicode channel.}\label{dicode961}
\end{figure}

\begin{figure}[!t]
    \centering
    \includegraphics[scale=0.9]{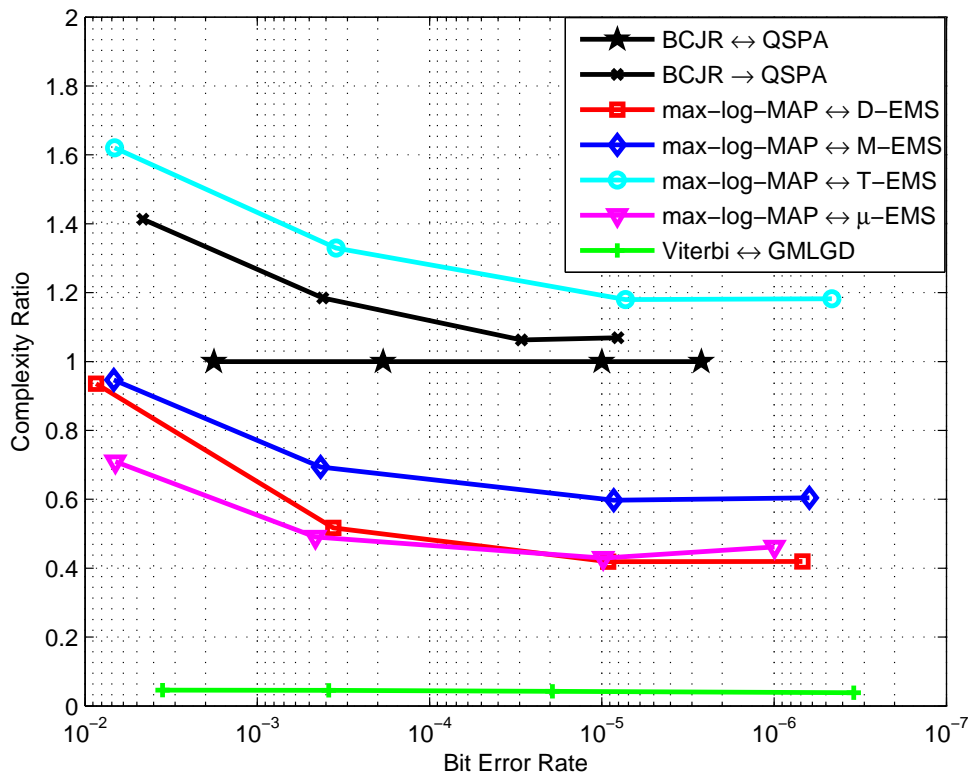}
    \caption{Complexity ratios of different detection/decoding algorithms for decoding the $\mathcal{C}_{32}[961,765]$ coded dicode channel.}
    \label{Complexity961765}
\end{figure}

{\bf Example 1:}~Consider the dicode channel with characteristic polynomial $f(D) = 1 - D $. The simulated code is the 32-ary LDPC code $\mathcal{C}_{32}[961,765]$ of rate 0.79, which is constructed by the properties of finite fields~\cite{Zeng08}. The corresponding parity-check matrix has row weight $30$ and column weights $10$ and $11$.
The squared Euclidean distances in {\bf Algorithm 1} are quantized with $ p = 9$ and $d_{max} = 80$. All the algorithms are carried out with maximum iteration $\mathcal{L} = 50$. The parameters of the max-log-MAP$\leftrightarrow$$X$-EMS algorithms are listed in the following:
    \begin{enumerate}
      \item  for the $\mu$-EMS algorithm, $\mu$ is calculated by~(\ref{mu_equation}) with $c = 1$; for the $D$-EMS algorithm, $D_s=50$, $D_b=40$; for the $T$-EMS algorithm, $T_s = 20$, $T_b = 10$; for $M$-EMS algorithm, $M =16$;
      \item the scaling factors of the $M$-EMS algorithm, the $T$-EMS algorithm, the $D$-EMS algorithm and the $\mu$-EMS algorithm are $0.4$, $0.4$, $0.3$ and $0.4$, respectively.
    \end{enumerate}
The simulation results are shown in the Fig.~\ref{dicode961}. It can be seen that at bit error rate~(BER) $10^{-5}$
\begin{enumerate}
      \item[a)] the max-log-MAP$\leftrightarrow$$X$-EMS algorithms perform as well as the BCJR$\leftrightarrow$QSPA;
      \item[b)] the max-log-MAP$\rightarrow$$X$-EMS algorithms have almost the same performances and suffer from
      performance degradations about 0.1 dB compared with BCJR$\leftrightarrow$QSPA;
      \item[c)] the Viterbi$\leftrightarrow$GMLGD algorithm suffers from performance degradations about 0.6 dB compared
      with BCJR$\leftrightarrow$QSPA.
\end{enumerate}
The complexity ratios of different detection/decoding algorithms are shown in Fig.~\ref{Complexity961765}. It can be
seen that, at BER=$10^{-5}$, the Viterbi$\leftrightarrow$GMLGD
algorithm is the simplest one with complexity ratio about 0.05, the max-log-MAP$\leftrightarrow$$\mu$-EMS and
the max-log-MAP$\leftrightarrow$D-EMS have almost the same complexity with complexity ratio 0.5. We also notice that both max-log-MAP$\leftrightarrow$$T$-EMS algorithm and BCJR$\rightarrow$QSPA are more complex than BCJR$\leftrightarrow$QSPA. This is because the complexity reduction per iteration of these two algorithms is not enough to counteract the complexity increase caused by the extra iterations\footnote{Actually, all other algorithms require more iterations than BCJR$\leftrightarrow$QSPA to converge in our simulations.}. In particular, $T$-EMS algorithm with a fixed performance-guaranteed threshold $T$ can not reduce too much computations at each iteration due to the large dynamic range of the messages. This motivated us to propose the $\mu$-EMS algorithm, which is similar to the $T$-EMS algorithm but with a dynamic and message-matched threshold.

%as the dynamic range of the
%value of the messages is large, a fixed $T$ which is independent of data is ineffective. This makes the
%max-log-MAP$\leftrightarrow$$T$-EMS algorithm requires more operations than BCJR$\leftrightarrow$QSPA. Note that BCJR$\rightarrow$QSPA requires more
%operations than BCJR$\leftrightarrow$QSPA as more iterations are required by BCJR$\rightarrow$QSPA to converge.

\begin{figure}[!t]
\centering
\includegraphics[scale=0.8]{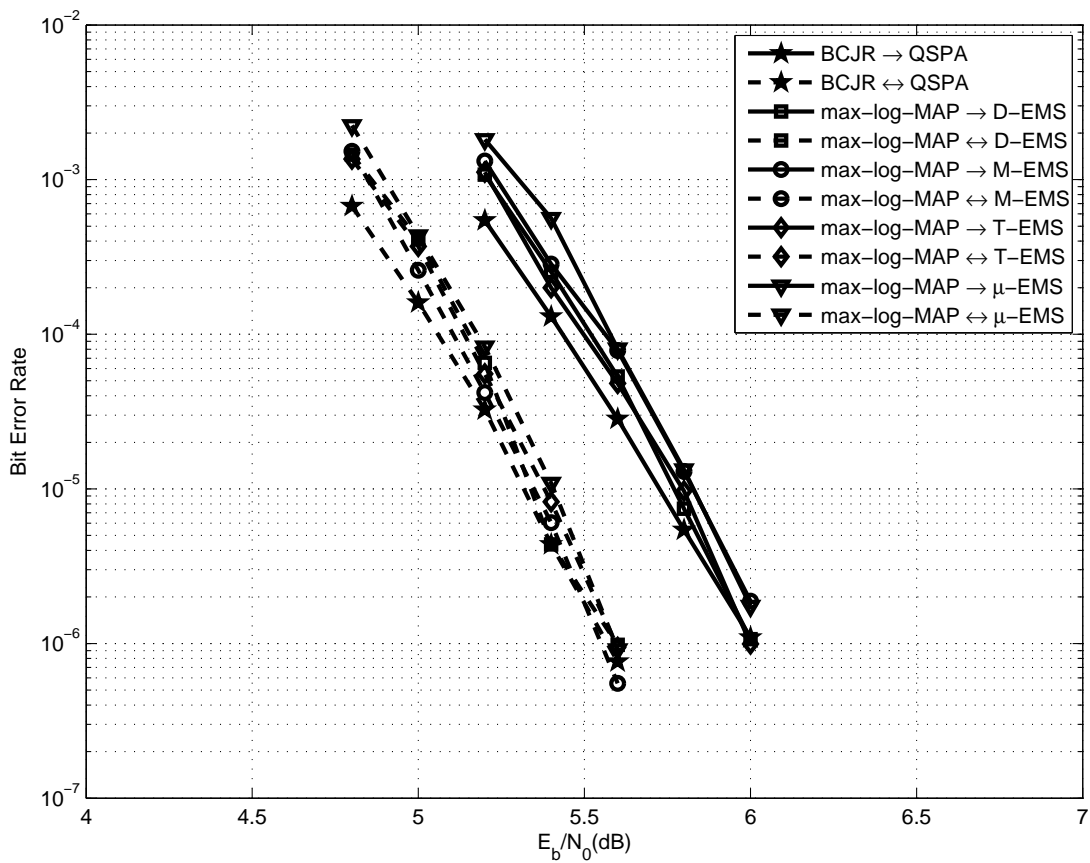}
\caption{Error performances of different detection/decoding algorithms for decoding the $\mathcal{C}_{16}[225,173]$ coded EPR4 channel.}\label{Epr4_225}
\end{figure}

{\bf Example 2:}~Consider an EPR4 channel with characteristic polynomial $f(D) = 1 + D - D^2 - D^3$. The simulated code is a 16-ary NB-LDPC code $\mathcal{C}_{16}[225,173]$ of rate 0.77, which is constructed by the properties of finite fields~\cite{Zeng08}.
The corresponding parity-check matrix has row weight $14$ and column weights $3$ and $4$. The squared Euclidean distances in
{\bf Algorithm 1} are quantized with $ p = 9$ and $d_{max} = 180$. All the algorithms are carried out with maximum iteration
$\mathcal{L} = 50$. The parameters for simulation are listed in the following:
    \begin{enumerate}
      \item for the $\mu$-EMS algorithm, $\mu$ calculated by~(\ref{mu_equation}) with $c = 0$; for the $D$-EMS algorithm, $D_s=45$, $D_b=35$; for the $T$-EMS algorithm, $T_s = 30$, $T_b = 10$; for $M$-EMS algorithm, $M =10$;
      \item the scaling factors of the $M$-EMS algorithm, the $T$-EMS algorithm, the $D$-EMS algorithm and the $\mu$-EMS
      algorithm are $0.6$, $0.6$, $0.6$ and $0.75$, respectively.
    \end{enumerate}
The simulation results are shown in the Fig.~\ref{Epr4_225}. It can be seen that at BER=$10^{-5}$
\begin{enumerate}
      \item[a)] the max-log-MAP$\leftrightarrow$$X$-EMS algorithms perform as well as the BCJR$\leftrightarrow$QSPA;
      \item[b)] the max-log-MAP$\rightarrow$$X$-EMS algorithms perform as well as the BCJR$\rightarrow$QSPA;
      \item[c)] the max-log-MAP$\leftrightarrow$$X$-EMS algorithms perform about 0.4 dB better than BCJR$\rightarrow$QSPA.
\end{enumerate}
\begin{figure}[!t]
\centering
\includegraphics[scale=0.8]{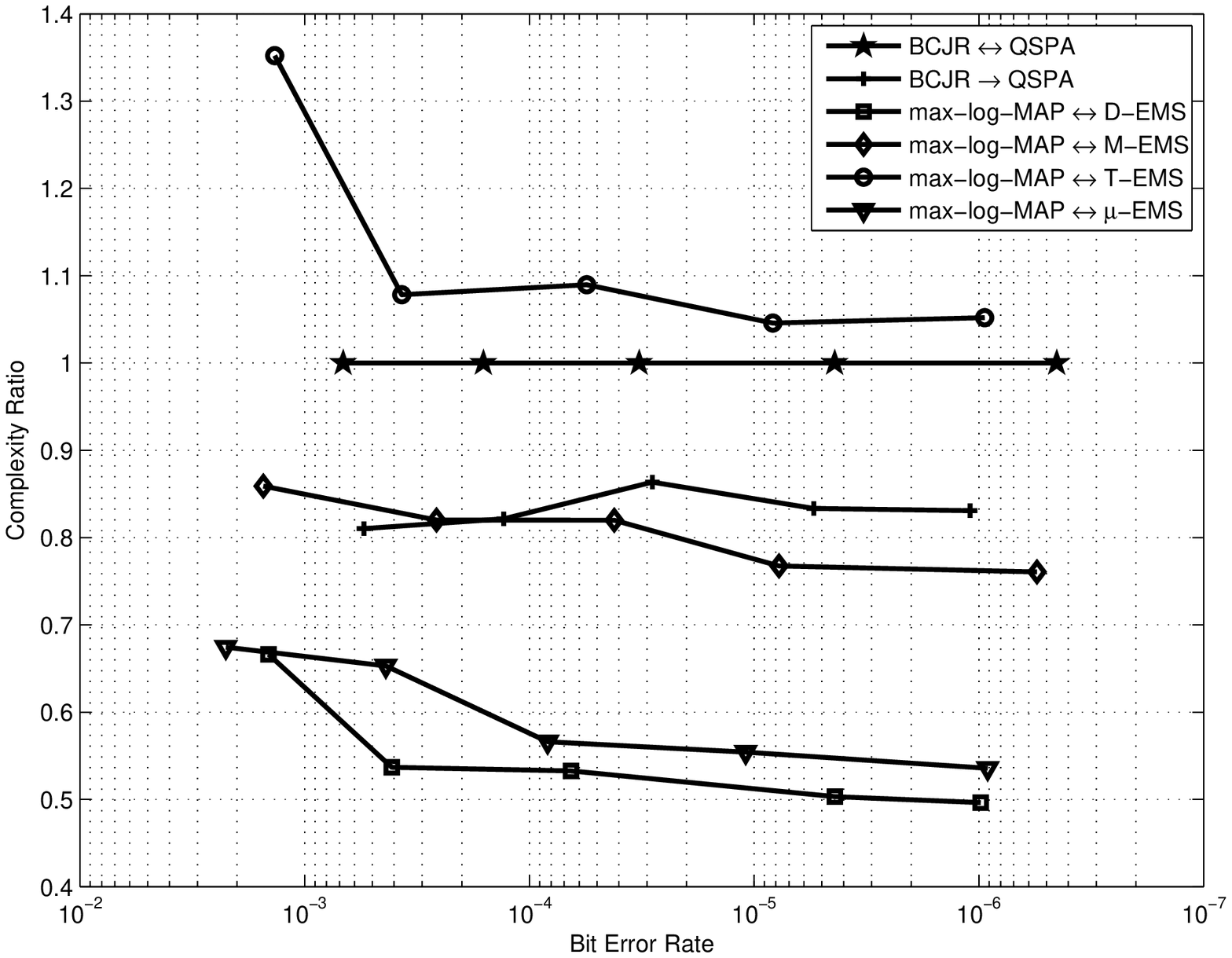}
\caption{Complexity ratios of different detection/decoding algorithms for decoding the $\mathcal{C}_{16}[225,173]$ coded EPR4 channel.}\label{complexity_Epr4}
\end{figure}
The complexity ratios of different detection/decoding algorithms are shown in Fig.~\ref{complexity_Epr4}. It can be
seen that, at BER=$10^{-5}$, the max-log-MAP$\leftrightarrow$D-EMS algorithm is the simplest one with complexity ratio
about 0.5, the max-log-MAP$\leftrightarrow$$\mu$-EMS algorithm has a complexity with complexity ratio about 0.55.
\begin{figure}[!t]
\centering
\includegraphics[scale=0.8]{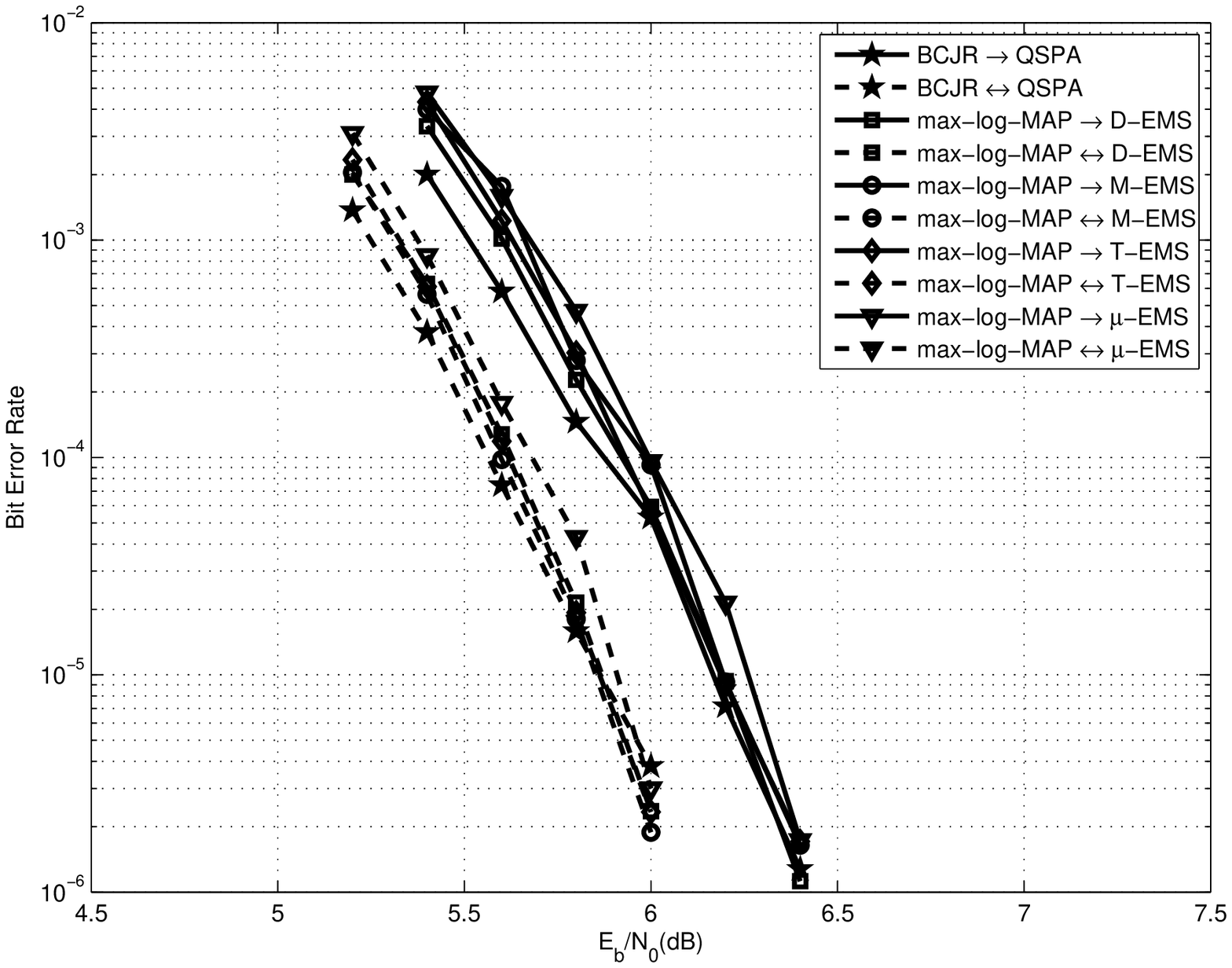}
\caption{Error performances of different detection/decoding algorithms for decoding the $\mathcal{C}_{16}[225,173]$ coded  Proakis.~(b) channel.}\label{hh_0507}
\end{figure}

{\bf Example 3:}~Consider the Proakis.~(b) channel~\cite[Sec~9.4-3]{Proakis} with characteristic polynomial $f(D) = 0.407 + 0.815D +0.407D^2 $. The simulated code is a 16-ary NB-LDPC code $\mathcal{C}_{16}[225,173]$ of rate 0.77, which is constructed by the properties of finite fields. The corresponding parity-check matrix has row weight $14$ and column weights $3$ and $4$. The squared Euclidean distances in~{\bf Algorithm 1} are quantized with $ p = 9$ and $d_{max} = 60$. All the algorithms are carried out with maximum iteration $\mathcal{L} = 50$. The parameters for simulation are listed in the following:
    \begin{enumerate}
        \item for the $\mu$-EMS algorithm, $\mu$ calculated by~(\ref{mu_equation}) with $c = 0$; for the $D$-EMS algorithm, $D_s=45$, $D_b=35$; for
         the $T$-EMS algorithm, $T_s = 10$, $T_b = 5$; for $M$-EMS algorithm, $M =10$;
        \item the scaling factors of the $M$-EMS algorithm, the $T$-EMS algorithm, the $D$-EMS algorithm and the $\mu$-EMS
         algorithm are $0.7$, $0.6$, $0.6$ and $0.75$, respectively.
   \end{enumerate}
\begin{figure}[!t]
\centering
\includegraphics[scale=0.8]{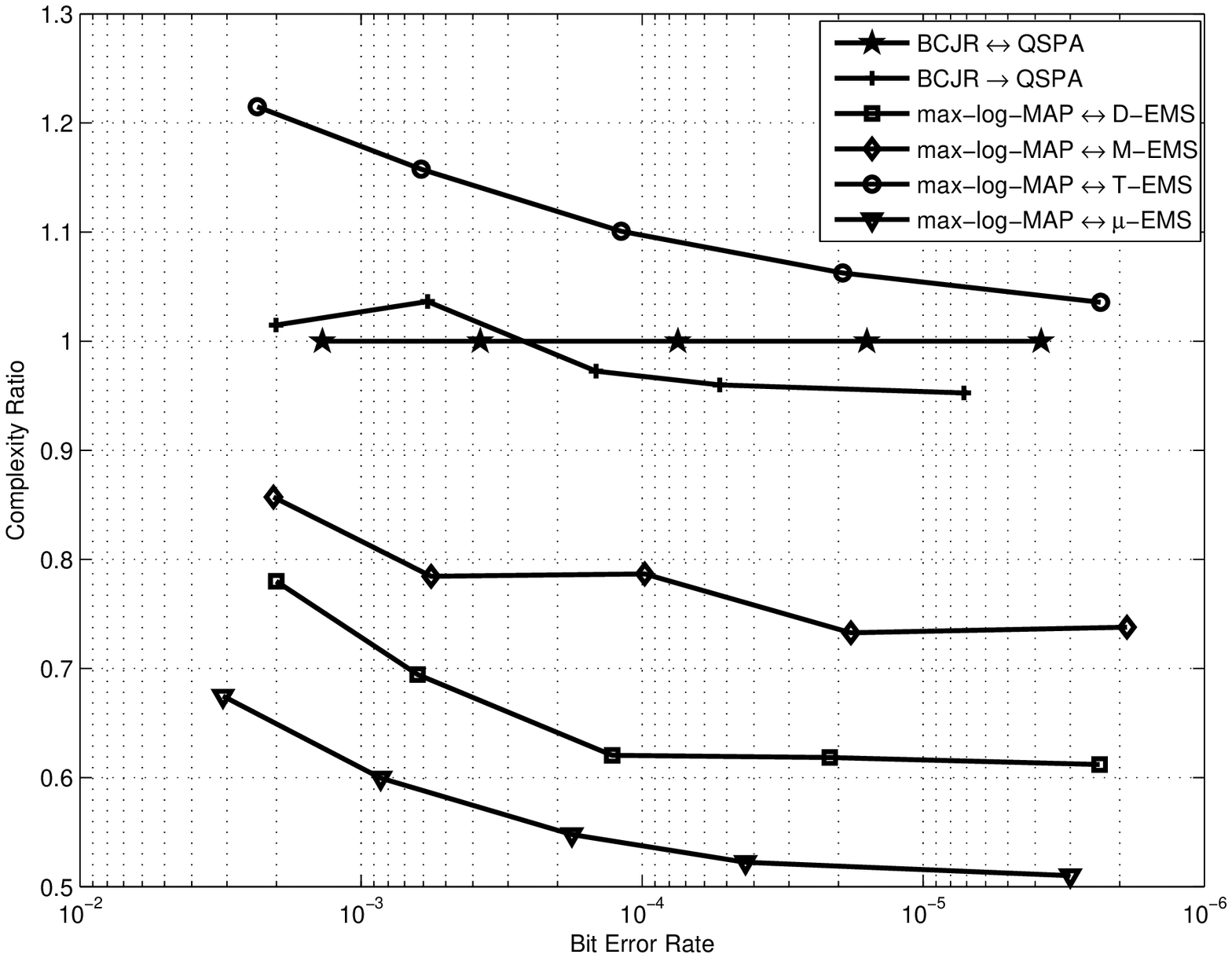}
\caption{Complexity ratios of different detection/decoding algorithms for decoding the $\mathcal{C}_{16}[225,173]$ coded  Proakis.~(b) channel.}\label{complexity_hh}
\end{figure}
The simulation results are shown in the the Fig.~\ref{hh_0507}. It can be seen that at BER=$10^{-5}$
\begin{enumerate}
      \item[a)] the max-log-MAP$\leftrightarrow$$X$-EMS algorithms perform as well as the BCJR$\leftrightarrow$QSPA;
      \item[b)] the max-log-MAP$\rightarrow$$X$-EMS algorithms perform as well as the BCJR$\rightarrow$QSPA;
      \item[c)] the max-log-MAP$\leftrightarrow$$X$-EMS algorithms perform about 0.3 dB better than BCJR$\rightarrow$QSPA.
\end{enumerate}

The complexity ratios of different detection/decoding algorithms are shown in Fig.~\ref{complexity_hh}. It can be
seen that, at BER=$10^{-5}$, the max-log-MAP$\leftrightarrow$$\mu$-EMS algorithm is the simplest one with complexity ratio
about 0.5.

{\bf Remark:~}From the preceding examples, it can be seen that the complexity ratio of max-log-MAP$\leftrightarrow$$\mu$-EMS algorithm
is always around 0.5.

\section{Conclusion}\label{conclusion}
In this paper, we have proposed two low-complexity joint iterative detection/decoding algorithms for NB-LDPC coded ISI channels.
The proposed algorithms work iteratively by exchanging either soft or hard messages between the detectors and the decoders. We have also presented a low-complexity decoding algorithm NB-LDPC codes. Simulation results show that the max-log-MAP$\leftrightarrow$$X$-EMS algorithm performs as well as
BCJR$\leftrightarrow$QSPA, and the Viterbi$\leftrightarrow$GMLGD algorithm, which is the simplest one, suffers from a performance degradation
within one dB compared with BCJR$\leftrightarrow$QSPA. These algorithms provide good candidates for trade-offs between performance and complexity.

%The iterative joint decoding algorithm employs two detectors/decoders at the destination to recover the transmitted data $\underline{u}^{t}$ from two successive received signals $\underline{y}^{t}_D$ and $\underline{y}^{t+1}_D$ in an iterative manner resembling the ``turbo equalization" as mentioned in~\cite{Khojastepour04}~\cite{Jun07}.

%\section{Complexity analysis and simulation results}\label{g}

%\subsection{Complexity analysis}\label{complexity}

%\subsection{Simulation results}

%\section{Conclusion}\label{conclusion}

% or


\appendix
\section{}
\begin{figure}[!t]
\centering
\includegraphics[scale=0.85]{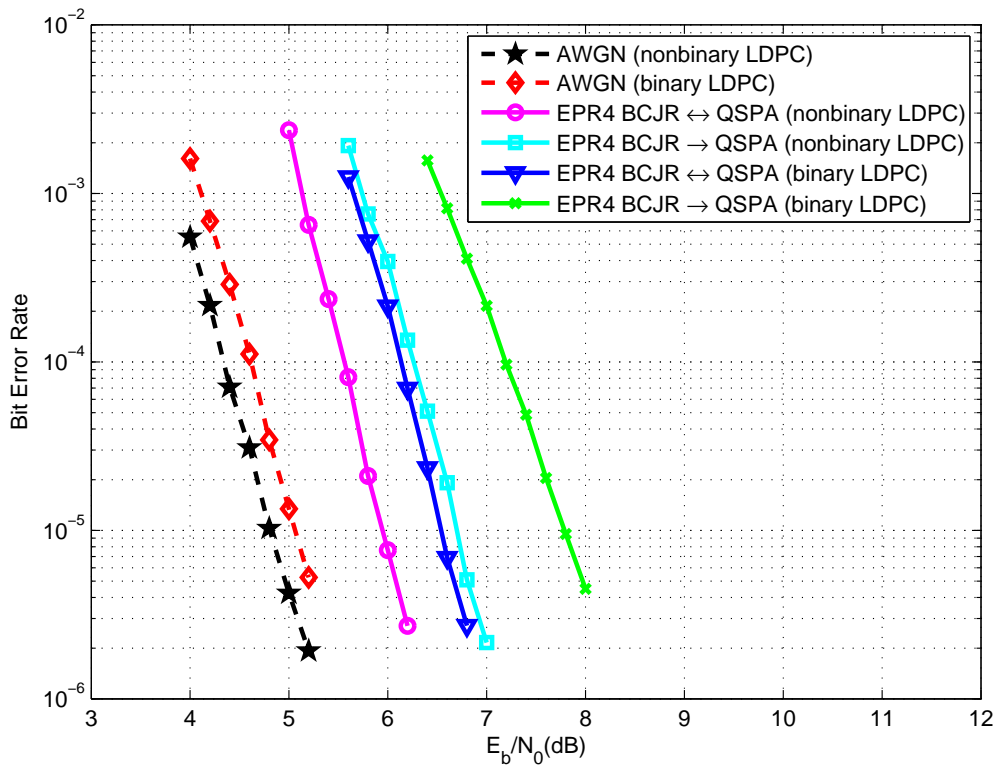}
\caption{Error performances of different algorithms for decoding the $\mathcal{C}_{16}[124,107]$ and binary LDPC code (62,495) coded AWGN channel and the EPR4 channel.}\label{awgn}
\end{figure}

A rough comparison between binary and NB-LDPC codes coded ISI channel is conducted in this appendix. We have simulated a binary LDPC code $\mathcal{C}_{2}[495,433]$~\cite{MacKayWeb} and a 16-ary NB-LDPC code $\mathcal{C}_{16}[124,107]$. These two codes have almost the same bit lengths and code rates. The nonbinary code $\mathcal{C}_{16}[124,107]$ is constructed by the PEG algorithm~\cite{Hu05} with column wight 2. We have simulated these two codes over EPR4 channels. The simulation results are shown in~Fig.~\ref{awgn}. It can be seen that $\mathcal{C}_{16}[124,107]$ performs about 0.6 dB better than $\mathcal{C}_{2}[495,433]$. We have also simulated these two codes over AWGN channels. The simulation results are also given in Fig.~\ref{awgn}. It can be seen that $\mathcal{C}_{16}[124,107]$ performs only 0.2 dB better than $\mathcal{C}_{2}[495,433]$ as apposed to 0.6 dB. We conclude that NB-LDPC codes may be more suitable to combat inter-symbol interferences.

%The performances of $\mathcal{C}_{16}[124,107]$ coded EPR4 channels when decoded with our algorithms are shown in~Fig.~\ref{epr4_78}. The squared Euclidean distances in {\bf Algorithm I} are quantized with $ p = 9$ and $d_{max} = 180$. All the algorithms are carried out with maximum iteration $\mathcal{L} = 50$. The parameters of the max-log-MAP$\leftrightarrow$$X$-EMS algorithms are listed in the following:
%    \begin{enumerate}
%      \item  for the $\mu$-EMS algorithm, $\mu$ calculated by~(\ref{mu_equation}) with $c = 1$; for the $D$-EMS algorithm, $D_s=50$, $D_b=40$; for the $T$-EMS algorithm, $T_s = 20$, $T_b = 10$; for $M$-EMS algorithm, $M =10$;
%      \item the scaling factors of the $M$-EMS algorithm, the $T$-EMS algorithm, the $D$-EMS algorithm and the $\mu$-EMS algorithm are $0.8$, $0.7$, $0.7$ and $0.7$, respectively.
%    \end{enumerate}
%It can be seen that, at bit error rate~(BER) $10^{-5}$, the max-log-MAP$\leftrightarrow X$-EMS algorithms have almost the same performances as BCJR$\leftrightarrow$QSPA.
%\begin{figure}[!t]
%% Requires \usepackage{graphicx}
%\centering
%\includegraphics[scale=0.8]{epr4_78}
%\caption{Error performances of different detection/decoding algorithms for decoding the $\mathcal{C}_{16}[124,107]$ coded EPR4 channel.}\label{epr4_78}
%\end{figure}

%\section{}

\section*{Acknowledgment}
The authors wish to thank Dr. Haiqiang Chen from Guangxi University for helpful comments to improve the presentation of this paper.

\ifCLASSOPTIONcaptionsoff

\newpage
\fi



%\bibliographystyle{IEEEtran}

%\bibliographystyle{ieeetr}
%\bibliography{tzzt}

%\makeatletter
%\def\@biblabel#1{}
%\makeatother
%\section{}

\bibliographystyle{ieeetr}
\bibliography{XEMS_IET}

%\end{spacing}
\end{document}